\documentclass[useAMS,usenatbib,referee]{mn2e}
\usepackage{graphicx}
\usepackage{epstopdf}
\usepackage[figuresright]{rotating}
\usepackage{subfigure,color}
\usepackage{longtable}
\usepackage{xspace}

\newcommand{\ion}[2]{{\textrm{#1}}\,{\textrm{\sc #2}}}

\title[Metallicity evolution of AGNs from UV emission-lines based on a new index]{Metallicity evolution of AGNs from UV emission-lines 
based on a new index}
\author[Dors et al.]
{Oli L. Dors Jr.$^{1}$\thanks{E-mail:olidors@univap.br}, M\'onica V.\ Cardaci$^{2,3}$,
Guillermo F. H\"agele$^{2,3}$,   \^Angela C. Krabbe$^1$\\
$^1$ Universidade do Vale do Para\'iba, Av. Shishima Hifumi, 2911, Cep
12244-000, S\~ao Jos\'e dos Campos, SP, Brazil\\
 $^2$ Consejo Nacional de Investigaciones Cient\'ificas y T\'ecnicas (CONICET), Argentina.\\
$^3$Facultad de Ciencias Astron\'omicas y Geof\'{\i}sicas, Universidad Nacional de La Plata, Paseo del Bosque s/n, 1900 La Plata, Argentina.\\}

\begin{document}

\date{Accepted- 2011 April 28. Received -2011 February 18.}

\pagerange{\pageref{firstpage}--\pageref{lastpage}} \pubyear{2011}

\maketitle

\label{firstpage}

\begin{abstract}

We analyzed the evolution of the metallicity of the gas with the redshift
for a sample of AGNs in a very wide redshift range ($0<z<4$) using
ultraviolet emission-lines from the narrow-line regions (NLRs) and
photoionization models. The  new index
C43=log[(\ion{C}{iv}+\ion{C}{iii}])/\ion{He}{ii}] is suggested as a
metallicity indicator for AGNs. Based on this indicator,  we confirmed the
no metallicity evolution of NLRs with the redshift  pointed out by previous 
works.   We found that metallicity   of AGNs  shows  similar evolution   than
the one predicted by cosmic  semi-analytic models of galaxy formation set within
the Cold Dark Matter  merging hierarchy (for $z \la 3$). 
Our results predict a mean metallicity for local objects in agreement
with the solar value (12+log(O/H)=8.69). 
This value is about the same that the maximum oxygen abundance value derived
for the central parts of   local  spiral galaxies.     Very low metallicity
$\log(Z/Z_{\odot})\approx -0.8$  for some objects in the range  $1.5 \:< \:z \:<
\:3$ is derived. 
  \end{abstract}



\section{Introduction}

The study of the metallicity in galaxies  and the knowledge of the chemical evolution of these
objects  with the redshift play an important role   to understand the formation and
evolution of the universe.   

 In general, models of cosmic chemical evolution  predict that the galaxy metallicities  
increase with the  aging of the universe. For example, \citet{malaney96}, using neutral hydrogen density obtained from
observations of Damped Lyman Alpha objects (DLAs) and an analytic model, showed that, for redshift
($z$) from about 4 to 0,  the metallicity  ($Z$)
rises  from  $0.05Z_{\odot}$ to 0.6\,$Z_{\odot}$. Other models, such as the  model of
\citet{pei99}, predict a  steeper increase of $Z$ with the cosmological time-scale. 
  From an observational point of view, the relation between the metallicity and the redshift, $Z-z$ relation, is
controversial. Along decades,  metallicity determinations 
of DLAs, using mainly the absorption line of the Zn (e.g. \citealt{pettini94}), 
 have been used to test  cosmic chemical evolution models (e.g. \citealt{kulkarni13, battisti12, somerville01, pei95}).
 Despite the large scattering  in the metallicity  for a fixed redshift, it has been confirmed the increase of the $Z$ with the time
 (e.g. \citealt{rafelski12}).  The same result is also found by observational studies of the gas phase metallicity of star-forming galaxies
   (e.g. \citealt{maiolino08, savaglio05})   and by metallicity
   studies of Narrow-Line Regions (NLRs) of high-$z$ radio 
   galaxies \citep{debreuck00}.  However, opposite results have also been
   obtained. For example,   
   \citet{mannucci10} from spectroscopic data of star-forming galaxies showed that there is a significant dependence of the
   gas-phase metallicity on the star-formation rate which, if taken into account, does not yield metallicity evolution
   with the redshift, at least for $z\:<2$.

  Moreover,  some studies based on emission-lines from  active galaxies have  failed 
   to identify the cosmic chemical evolution. For example, \citet{dietrich03a}  compared  rest-frame  of     broad emission-line  
  intensities in the  ultraviolet  of  a sample of 70 quasars ($z  \ga 3.5$) with photoionization models
  results of \citet{hamann02}.  They  found that the  objects analyzed have  an  average  metallicity of  about  4-5 $Z_{\odot}$, which 
  is in disagreement with the $Z$ determinations using absorption lines (see \citealt{battisti12, kulkarni05}).
   A similar analysis performed by \citet{nagao06a} using ultraviolet spectra of NLRs for objects with  redshifts between 1.2 and 4.0
pointed out a constant behavior of the gas metallicity with $z$.  Nagao and
collaborators  interpreted the lack of evolution of $Z$ obtained from NLRs as a result of the fact that 
the major epoch of star formation in the host galaxies of active nuclei  is at very high redshifts ($z\ga4$). 
 Also \citet{matsuoka09} obtained   UV rest-frame  spectral data from  the narrow-line region  of 9 high-z radio galaxies at $z > 2.7$  and, combining
these with data from the literature, found not significant metallicity evolution in NLRs  for 
$z \la 4$.

 Metallicity indicators based on emission-line ratios can be subject to uncertainties (e.g. \citealt{dors11}).  
 In fact, the \ion{N}{v}\,$\lambda$1240/\ion{C}{iv}\,$\lambda$1549  ratio, generally
 used as metallicity indicator for AGNs \citep{hamann92}, can yield $Z$ estimations somewhat  uncertain since the 
 $\ion{N}{v}$ emission  line could be enhanced by Ly$\alpha$ photons scattered in a broad absorption-line  wind  
 (see \citealt{hamann02} and references therein). Moreover, any metallicity indicator based on nitrogen-lines must take into 
 account a N/O abundance relation with the metallicity \citep{perez09}, which is  poorly   determined for AGNs.
  In this sense, metallicity indicators based on carbon emission-lines, such as the \ion{C}{iv}$\lambda$1549/\ion{He}{ii}$\lambda$1640
  suggested by \citet{nagao06a}, can be more reliable. Although the relation between
   C/O abundance ratio and the O/H (used as metallicity tracer) must to be taken into account  in calibrations \citep{garnett04},   
 chemical evolution models of QSOs of \citet{hamann93} predict a C/O abundance ratio nearly constant for
 objects chemically evolved, i.e older than 1 Gyr. This does decrease the uncertainties in metallicity determinations  
 based on carbon emission-lines.

In this paper, we report an analysis of the chemical evolution
of AGNs with the cosmological time-scale by modelling the 
ultraviolet narrow emission-lines observed at different  redshifts. 
   We proposed a new metallicity indicator calibrated taking also into
  account its dependence on other parameters than the
metallicity. The paper is organized as follows. In  Section~\ref{obs} 
we describe the observational data used along the paper.  
A description of the photoionization models used  in the paper
  is given in Sect.~\ref{mod}.  In Sect.~\ref{calib} 
a new metallicity tracer is presented. The results of the use of this index
and the discussion are presented in Sects.~\ref{res} and \ref{disc}, respectively. The
final conclusions is given in  Sect.~\ref{conc}.


\section{Observational data}
\label{obs}

 The fluxes of the  \ion{N}{v}$\lambda$1240, \ion{C}{iv}$\lambda$1549,
\ion{He}{ii$\lambda$}1640, and \ion{C}{iii}]$\lambda$1909  emission-lines originated in the NLRs
of a sample of Seyfert 2 (12 objects), high-$z$ radio galaxies (59 objects)  and   type 2 quasars (10 objects) with redshifts $0 \la z \la 4.0$ were compiled from the literature. 
The sample is about the same that the one compiled by \citet{nagao06a}  with the addition of  Seyfert 2  data 
 taken from \citet{kraemer94} and    from \citet{diaz88}.   In Table~\ref{tab0} the identification, 
 redshift, adopted emission-line intensities, and the bibliographic
 reference of each considered object are presented.  The objects in this
   table are grouped by their nature. We did not consider in our sample the lines with only
   intensity upper limits reported.

\begin{table*}
\caption{Fluxes of emission-lines compiled from the literature.}
\vspace{0.3cm}
\label{tab0}
\begin{tabular}{|lccccccc}
\hline
                                   \multicolumn{8}{|c|}{Seyfert 2}  \\			 
\noalign{\smallskip}  
 Object                     &       redshift &      \ion{N}{v}$\lambda$1239           &     \ion{C}{iv}$\lambda$1549            &            \ion{He}{ii}$\lambda$1640         &       \ion{C}{iii}]$\lambda$1909          &  Flux units ($\rm erg/s/ cm^{2}$)  &Reference          \\
NGC\,1068               &        0.004   &         224$\pm41$                           &          520$\pm80$                          &           187$\pm29$               &             240$\pm35$                        &  $10^{-14}$                               &  1    \\
NGC\,4507		&        0.012   &          5.2$\pm1.0$                         &          13.5$\pm2.7$                         &            5.6$\pm1.1$              &             5.8$\pm1.2$                        & $10^{-14}$                                &  1    \\
NGC\,5135		&        0.014   &          1.1$\pm0.2$                         &           4.1$\pm0.8$                          &           10.0$\pm2.0$             &             ---                                       & $10^{-14}$                                &  1      \\
NGC\,5506		&        0.006   &         ---                                         &           4.5$\pm1.4$                          &            2.0$\pm0.6$              &             3.6$\pm0.7$                        & $10^{-14}$                                &  1     \\
NGC\,7674		&        0.029   &         ---                                         &           11.4$\pm3.3$                        &             5.1$\pm1.5$             &             7.9$\pm2.7$                        & $10^{-14}$                                &   1    \\
Mrk\,3                 	    &        0.014   &          3.0$\pm1.0$                         &            21$\pm2$                             &            9$\pm1$                   &             9$\pm1$                              & $10^{-14}$                                &   1    \\		
Mrk\,573        	  &        0.017    &        6.3$\pm0.9$                          &            29$\pm4.3$                          &          12.6$\pm1.9$              &            8.8$\pm1.3$                          &  $10^{-14}$                                &   1      \\		
Mrk\,1388	         &        0.021    &        ---                                          &           8.3$\pm1.2$                         &            3.8$\pm0.6$              &            3.6$\pm0.5$                          & $10^{-14}$                                &    1    \\	
MCG-3-34-64	     &        0.017     &        5.0$\pm0.1$                          &            14$\pm3$                             &            10$\pm2$                  &            7$\pm1$                               &  $10^{-14}$                               &    1   \\	
NGC\,7674              &        0.029     &       ---                                          &         26$\pm1.40$                        &            10$\pm3$                 &           18.36$\pm6.21$                     &   $10^{-14}$                               &    2  \\
IZw\,92                   &        0.037     &       ---                                           &         9.7$\pm2.8$                           &          1.46$\pm0.43$           &            ---                                         & $10^{-13}$                                 &    2  \\  
NGC\,3393             &        0.012      &      1.15                                         &        47.75                                      &           25.73                        &            ---                                         & $10^{-14}$                                 &   3  \\
\hline
                                   \multicolumn{8}{|c|}{ Type 2 Quasar}  \\			 
CDFS-027                &     3.064       &       2.5$\pm0.7$                            &           6.4$\pm0.5$                         &             2.3$\pm0.9$           &           ---                                          &  $10^{-18}$                               &  1    \\
CDFS-031                &     1.603       &      ---                                            &           24.1$\pm1.4$                       &            13.3$\pm1.2$          &           10.3$\pm1.3$                          &  $10^{-18}$                               &  1    \\
CDFS-057                &     2.562       &       8.4$\pm1.4$                            &           17.8$\pm0.8$                       &            7.6$\pm0.8$            &            13.3$\pm0.9$                         &  $10^{-18}$                               &  1    \\
CDFS-112a              &     2.940        &     14.6$\pm0.8$                           &           15.2$\pm1.0$                        &            8.9$\pm0.9$            &           4.5$\pm0.8$                            &  $10^{-18}$                               &  1    \\  
CDFS-153                &     1.536        &     ---                                           &           25.5$\pm1.4$                        &             6.2$\pm1.1$            &           13.7$\pm1.6$                          &  $10^{-18}$                               &  1    \\  
CDFS-202                &     3.700        &      26.8$\pm1.1$                          &           38.9$\pm1.2$                       &            19.7$\pm1.5$           &           ---                                           &  $10^{-18}$                               &  1    \\            
CDFS-263b              &     3.660        &      4.6$\pm0.7$                            &           15.5$\pm0.8$                       &             ---                           &           ---                                           &  $10^{-18}$                               &  1    \\  
CDFS-531                &     1.544        &      ---                                           &           22$\pm1.4$                          &            17.4$\pm1.5$           &           14.4$\pm1.5$                          &  $10^{-18}$                               &  1    \\    
CDFS-901                &     2.578        &      6.5$\pm0.8$                            &           19.7$\pm1.0$                        &             ---                           &            3.3$\pm0.9$                           &  $10^{-18}$                               &  1    \\    
CXO\,52                   &     3.288        &      6$\pm1.2$                               &           35$\pm2$                             &              17$\pm2$              &            21$\pm2$                               &  $10^{-18}$                               &  1    \\             
\hline
                                 \multicolumn{8}{|c|}{High-z radio galaxy}  \\			 
TN\,J0121+1320       &     3.517         &          ---                                        &     0.263$\pm0.005$                       &     0.330$\pm0.012$           &         0.282$\pm0.009$                       &     $10^{-16}$                            & 4     \\
TN\,J0205+2242       &     3.507         &          ---                                        &     0.873$\pm0.025$                       &     0.519$\pm0.046$           &         0.418$\pm0.049$                       &      $10^{-16}$                            & 4    \\  
MRC\,0316-257        &     3.130         &         ---                                         &     0.267$\pm0.011$                       &     0.301$\pm0.009$           &         0.345$\pm0.018$                       &      $10^{-16}$                            & 4    \\     
USS\,0417-181         &     2.773         &        ---                                          &     0.356$\pm0.026$                       &     0.492$\pm0.019$           &         0.553$\pm0.047$                       &      $10^{-16}$                            & 4    \\ 
TN\,J0920-0712        &     2.758         &        1.015$\pm0.014$                    &     3.365$\pm0.010$                       &     2.063$\pm0.011$           &         1.945$\pm0.028$                       &       $10^{-16}$                            & 4    \\ 
WN\,J1123+3141      &     3.221         &        1.698$\pm0.013$                    &     1.570$\pm0.011$                       &     0.425$\pm0.014$           &         0.183$\pm0.028$                       &       $10^{-16}$                            & 4    \\ 
4C\,24.28                 &     2.913         &        1.225$\pm0.012$                    &     1.235$\pm0.020$                       &     0.978$\pm0.011$           &          0.812$\pm0.041$                      &       $10^{-16}$                            & 4    \\ 
USS\,1545-234         &     2.751         &        1.335$\pm0.031$                    &     1.343$\pm0.021$                       &     0.878$\pm0.012$           &          0.606$\pm0.031$                      &       $10^{-16}$                            & 4    \\ 
USS\,2202+128        &      2.705         &       0.160$\pm0.019$                    &      0.704$\pm0.012$                      &     0.289$\pm0.010$            &         0.292$\pm0.011$                      &       $10^{-16}$                            & 4    \\ 
USS\,0003-19          &      1.541         &        ---                                          &     5.90                                          &      3.90                              &         3.40                                         &        $10^{-16}$                            & 5    \\ 
BRL\,0016-129         &      1.589         &       ---                                          &      1.60                                          &      ---                                 &       2.60                                          &        $10^{-16}$                            & 5    \\ 
MG\,0018+0940       &       1.586         &      ---                                           &      0.81                                         &    0.42                                &       0.87                                          &        $10^{-16}$                            & 5    \\ 
MG\,0046+1102       &       1.813         &     ---                                            &      0.65                                         &    0.55                                &       0.79                                          &         $10^{-16}$                            & 5    \\ 
MG\,0122+1923       &       1.595         &     ---                                            &      0.32                                         &    0.38                                &       0.32                                           &         $10^{-16}$                            & 5    \\ 
USS\,0200+015        &      2.229         &     ---                                            &      4.20                                         &     3.20                               &       4.00                                           &         $10^{-16}$                            & 5    \\
USS\,0211-122         &      2.336         &       4.10                                       &      5.60                                         &      3.10                               &      2.20                                           &         $10^{-16}$                            & 5    \\
USS\,0214+183        &      2.130         &       ---                                          &      3.00                                         &      1.80                               &      1.80                                           &         $10^{-16}$                            & 5    \\
MG\,0311+1532       &       1.986        &        ---                                          &      0.34                                         &      0.20                               &      0.21                                           &         $10^{-16}$                            & 5    \\ 
BRL\,0310-150         &       1.769        &       ---                                           &      10.20                                       &     4.00                                &      5.00                                            &         $10^{-16}$                            & 5    \\ 
USS\,0355-037        &        2.153        &      ---                                            &    2.70                                           &     3.70                                &     2.30                                            &         $10^{-16}$                            & 5    \\   
USS\,0448+091       &        2.037        &      ---                                            &    1.20                                           &     1.40                                &     2.70                                            &         $10^{-16}$                            & 5    \\  
USS\,0529-549        &        2.575        &      ---                                            &    0.40                                           &     0.60                                &     1.80                                             &         $10^{-16}$                            & 5    \\  
4C\,41.17                &        3.792        &     ---                                             &    1.32                                           &      0.55                               &      0.91                                             &         $10^{-16}$                            & 5    \\ 
USS\,0748+134       &        2.419        &     ---                                             &    1.80                                           &      1.50                              &     1.40                                              &         $10^{-16}$                            & 5    \\ 
USS\,0828+193       &        2.572        &    ---                                              &    1.90                                           &      1.90                               &     2.0                                               &           $10^{-16}$                            & 5    \\ 
4C\,12.32                &        2.468        &      ---                                            &    3.40                                           &     2.30                                 &    1.60                                             &           $10^{-16}$                            & 5    \\ 
TN\,J0941-1628       &        1.644        &     ---                                             &    3.20                                           &     0.90                                 &    2.00                                             &          $10^{-16}$                            & 5    \\ 
USS\,0943-242        &        2.923        &     1.70                                          &    3.90                                           &    2.70                                 &    2.30                                              &          $10^{-16}$                            & 5    \\ 
MG\,1019+0534       &        2.765        &     0.23                                          &    1.04                                           &    0.85                                 &   0.49                                               &          $10^{-16}$                            & 5    \\ 
TN\,J1033-1339       &        2.427        &     ---                                             &    2.30                                           &    0.80                                 &   0.70                                               &          $10^{-16}$                            & 5    \\ 
TN\,J1102-1651       &        2.111        &    ---                                              &    1.00                                           &    1.30                                 &   1.10                                               &          $10^{-16}$                            & 5    \\ 
USS\,1113-178        &        2.239        &   ---                                               &     1.70                                          &    0.70                                 &   2.80                                               &          $10^{-16}$                            & 5    \\ 
3C\,256.0                 &       1.824        &      1.40                                         &     5.23                                          &    5.47                                &    4.28                                              &          $10^{-16}$                            & 5    \\ 
USS\,1138-262         &       2.156        &      ---                                            &     0.80                                          &     1.30                                &    1.30                                              &          $10^{-16}$                            & 5    \\ 
BRL\,1140-114         &       1.935        &       ---                                            &    1.00                                          &    0.50                                 &    0.60                                              &          $10^{-16}$                            & 5    \\    
4C\,26.38                 &       2.608        &      ---                                             &    8.90                                          &    5.70                                 &    2.40                                              &          $10^{-16}$                            & 5    \\
MG\,1251+1104       &        2.322        &     ---                                              &    0.30                                         &    0.30                                 &     0.52                                              &          $10^{-16}$                            & 5    \\
WN\,J1338+3532     &        2.769         &    ---                                              &    1.30                                         &    3.00                                 &     2.20                                              &          $10^{-16}$                            & 5    \\
\hline
\end{tabular}
\end{table*}

\begin{table*}
\setcounter{table}{0}
\caption{-$continued$}
\vspace{0.3cm}
\begin{tabular}{|lccccccc}
\hline
                                   \multicolumn{8}{|c|}{High-z radio galaxy}  \\
\noalign{\smallskip}  
 Object                     &       redshift &      \ion{N}{v}$\lambda$1240              &     \ion{C}{iv}$\lambda$1549                    & \ion{He}{ii}$\lambda$1640     &       \ion{C}{iii}]$\lambda$1909      &  Flux units ($\rm erg/s/ cm^{2}$)  &Reference          \\

MG\,1401+0921       &       2.093          &  ---                                                &    0.41                                                   &    0.50                                  &    0.34                                   &          $10^{-16}$                            & 5    \\
3C\,294.0                 &      1.786          &  3.10                                             &    15.50                                                  &   15.50                                 &   18.60                                  &          $10^{-16}$                            & 5    \\
USS\,1410-001         &      2.363          &   1.68                                            &    3.36                                                    &   2.52                                   &   1.12                                   &          $10^{-16}$                            & 5    \\
USS\,1425-148         &      2.349          &   ---                                               &   2.30                                                     &  2.30                                    &   1.00                                   &          $10^{-16}$                            & 5    \\
USS\,1436+157        &      2.538           &  ---                                                &  17.0                                                      &  6.0                                      &  9.40                                   &          $10^{-16}$                            & 5    \\
3C\,324.0                 &      1.208           & ---                                                  &  3.67                                                     & 2.70                                     &  3.47                                   &          $10^{-16}$                            & 5    \\
USS\,1558-003         &      2.527           &   ---                                                &  2.70                                                     & 1.70                                     & 1.20                                    &          $10^{-16}$                            & 5    \\
BRL\,1602-174         &      2.043           &  ---                                                  &  10.0                                                     &  4.8                                     & 2.70                                    &          $10^{-16}$                            & 5    \\
TXS\,J1650+0955     &      2.510           &    ---                                                &  3.20                                                     &  2.70                                   & 1.20                                    &          $10^{-16}$                            & 5    \\
8C\,1803+661           &      1.610           &  ---                                                 &   5.30                                                     & 2.60                                    & 1.90                                    &          $10^{-16}$                            & 5    \\
4C\,40.36                 &       2.265           &  ---                                                &    6.20                                                     &5.60                                     &  5.90                                  &          $10^{-16}$                            & 5    \\
BRL\,1859-235         &       1.430            & ---                                                 &    3.40                                                    & 4.60                                    & 4.70                                   &          $10^{-16}$                            & 5    \\
4C\,48.48                 &       2.343            & ---                                                 &    6.10                                                    &  3.70                                   &  2.80                                  &          $10^{-16}$                            & 5    \\
MRC\,2025-218        &       2.630            & 0.62                                                &  0.69                                                      &  0.35                                  & 0.97                                   &          $10^{-16}$                            & 5    \\
TXS\,J2036+0256     &       2.130            &  ---                                                  &  0.60                                                      &  0.70                                  & 1.20                                   &          $10^{-16}$                            & 5    \\
MRC\,2104-242        &       2.491            &  ---                                                  &  3.80                                                      &  1.90                                  &  2.66                                  &          $10^{-16}$                            & 5    \\
4C\,23.56                 &       2.483            & 1.36                                                &  2.08                                                      &  1.52                                  & 1.28                                   &          $10^{-16}$                            & 5    \\
MG\,2121+1839        &      1.860             &  ---                                                  &  0.53                                                      &  0.14                                  &  0.24                                 &          $10^{-16}$                            & 5    \\
USS\,2251-089          &      1.986             & ---                                                  &  3.30                                                       &  1.30                                  & 1.50                                 &          $10^{-16}$                            & 5    \\
MG\,2308+0336         &      2.457             &  0.57                                              &  0.63                                                       & 0.39                                   & 0.45                                 &          $10^{-16}$                            & 5    \\
4C\,28.58                  &      2.891              &  ---                                                 & 0.30                                                       &  1.60                                   &  1.80                               &          $10^{-16}$                            & 5    \\
\hline
\end{tabular}
\begin{minipage}[c]{2\columnwidth}
References--- (1)  Data compiled by  \citet{nagao06a}, (2) \citet{kraemer94}, (3) \citet{diaz88}, (4) \citet{matsuoka09}, \\
(5) \citet{debreuck00}
\end{minipage}
\end{table*}

Since the emission-line intensities were not reddening
corrected  it could  yield some bias in our results. However,   
\citet{nagao06a}, using an extinction curve described by \cite{cardelli89}, showed
that the effect of dust extinction on the \ion{C}{iii}]/\ion{C}{iv}  and \ion{C}{iv}/\ion{He}{ii} emission-line ratios, generally used as 
ionization parameter and metallicity indicators of AGNs respectively, is not important. 
  It is worth to mention  that the data compiled from the literature were 
 obtained with different instrumentation and observational techniques. However, the effects caused by the use of
non-homogeneous  data, such as the ones used in this work, do not yield any bias on  the results
of abundance estimations in the gas phase of star-forming regions, as pointed out by  \citet{dors13}.

\begin{figure}
\centering
\includegraphics[angle=-90,width=1\columnwidth]{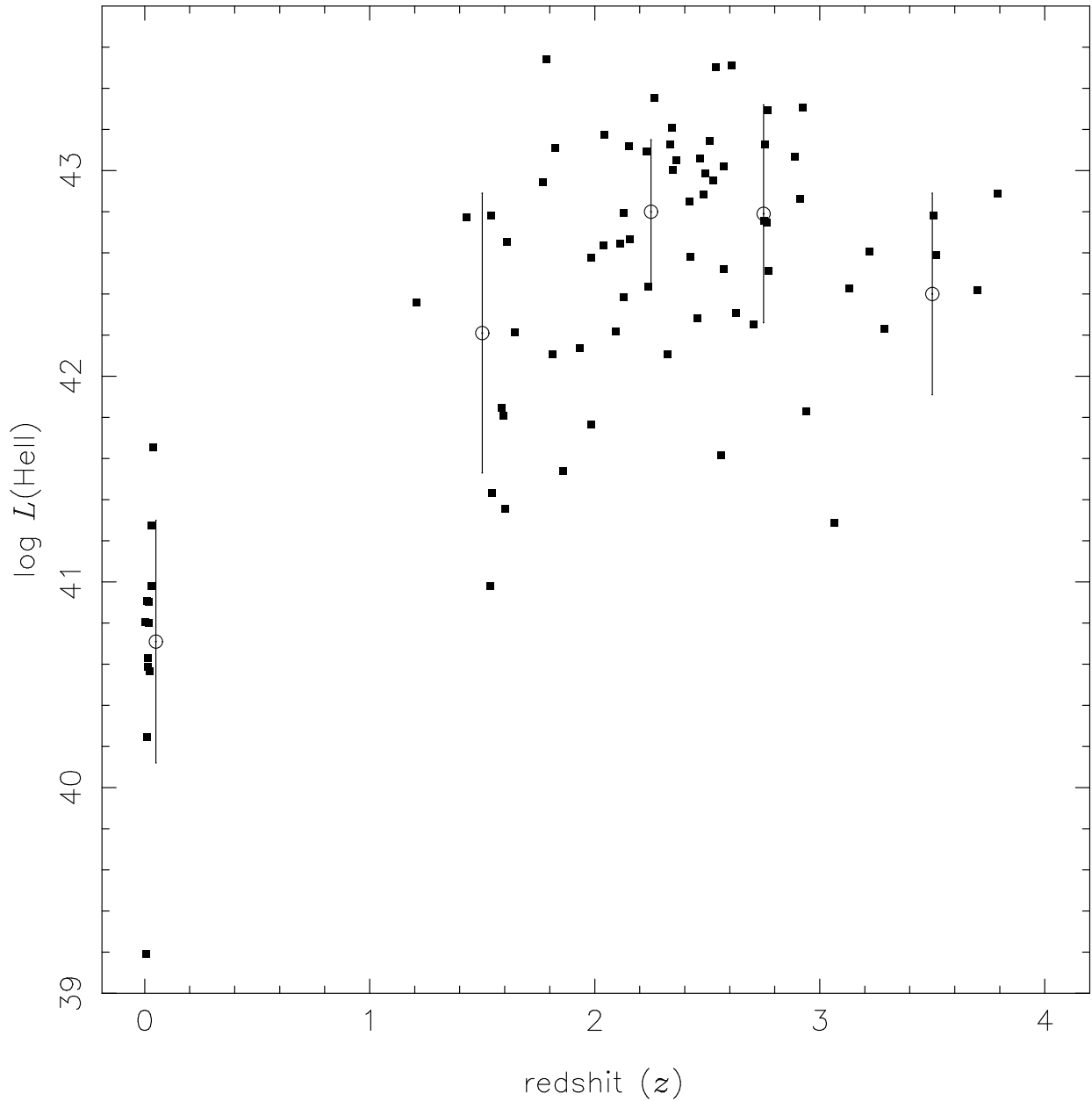}
\caption{Luminosity of \ion{He}{ii}$\lambda$1640   versus redshift. The squares
  represent the values  for the objects in Table~\ref{tab0}. 
The circles represent the  average and their error bars the standard
deviation of the  luminosity at each redshift interval (see Table~\ref{tab1}).}
\label{lumi}
\end{figure}

To investigate possible redshift evolutions of the AGN metallicity based on
heterogeneous sample, it is important to verify the effects of the dependence of the  metallicity
on the AGN luminosity, i.e the $Z-L$ relation (see \citealt{matsuoka09}).   For that, we used the  \ion{He}{ii}$\lambda$1640   luminosity ($L(\rm He\:II)$) as a
representative value for the bolometric luminosity, as suggested by \citet{matsuoka09}.
  The distance to each object
was calculated using the   $z$ value given in Table~\ref{tab0} and  assuming a spatially flat  cosmology with
 $H_{0}$\,=\,71 $ \rm km\:s^{-1} Mpc^{-1}$, $\Omega_{m}=0.270$,    and $\Omega_{\rm vac}=0.730$  \citep{wright06}.
In  Figure~\ref{lumi} we presented the values  of  $L(\rm He\:II)$ versus the
redshift for the objects in our sample. We computed the
average and the standard deviation of the  luminosity for 5 redshift intervals
and these values are given Table~\ref{tab1}  as well as the average values of the observed emission-lines
intensities for each interval of redshift considered.   
 We can note  the strong  dependence of the  $L(\rm He\:II)$ with the redshift, 
 probably due
to selection effects and that the intrinsic emission-line luminosity of nearby
Seyfert 2 galaxies is  
significantly smaller than that of the high-$z$ radio galaxies and type 2 quasars.  
Since more luminous AGNs have higher metallicity gas clouds (\citealt{matsuoka09, nagao06a}),  the $Z-L$ relation
must be taken into account in our analysis, in the sense that for high redshift we are analyzing a sample of most metallic
objects (more luminous).

\begin{table}
\centering
\caption{Logarithm  of the average values of $L(\rm He\:II)$, the observed UV emission-line intensity ratios for the selected redshift intervals, and
the  number of objects $N$ included in each interval.}
\label{tab1}
\vspace{0.2cm}
\begin{tabular}{@{}l rr rr r c @{}}
\hline 
 $z$        & $N$     &   $L(\rm He\:II)$ (erg/s)      &       \ion{C}{iv}/\ion{He}{ii}      &     \ion{C}{iii}]/\ion{C}{iv}      &  \multicolumn{1}{c}{\ion{N}{v}/\ion{He}{ii}}  &   $\frac{\ion{C}{iv}+\ion{C}{iii]}}{\ion{He}{ii}}$ \\
 \noalign{\smallskip} 
0-0.1      &   12    &        41.71$\pm0.59$                        &      $0.31\pm0.27$                 &    $-0.32\pm0.12$         &   $-0.47\pm0.51$                       &    $0.53\pm0.09$                                     \\       
1.0-2.0   &   18     &       42.21$\pm0.68$                        &      $0.24\pm0.22$                  &   $-0.14\pm0.19$         &   $-0.64\pm0.07$                        &   $0.46\pm0.17$                                      \\                                    
2.0-2.5   &   22     &        42.80$\pm0.35$                         &     $0.10\pm0.18$                  &    $-0.12\pm0.27$         &  $0.01\pm0.15$                         &    $0.37\pm0.15$                                     \\       
2.5-3.0   &   18     &       42.79$\pm0.53$                         &     $0.08\pm0.28$                  &    $-0.14\pm0.39$         & $-0.06\pm0.28$                         &     $0.37\pm0.17$                                    \\       
3.0-4.0   &    8    &       42.40$\pm0.49$                         &     $0.25\pm0.23$                  &     $-0.24\pm0.37$       &   $0.07\pm0.28$                        &    $0.44\pm0.16$                                                         \\       
\hline 
\end{tabular}
\end{table}

\section{Photoionization models}
 \label{mod}
 
 \subsection{Model parameters}
 
  In this paper a new metallicity indicator for AGNs is proposed. To obtain a calibration of
this indicator with the metallicity,    we 
built photoionization models using    Cloudy 08.00 \citep{ferland13}.
In these  models, predicted emission-line intensities depend basically on three parameters,
the spectral energy distribution (SED), the ionization parameter $U$ and the
metallicity $Z$.  In what follows the use of these parameters is discussed.

\begin{figure}
\centering
\includegraphics[angle=0,width=0.7\columnwidth]{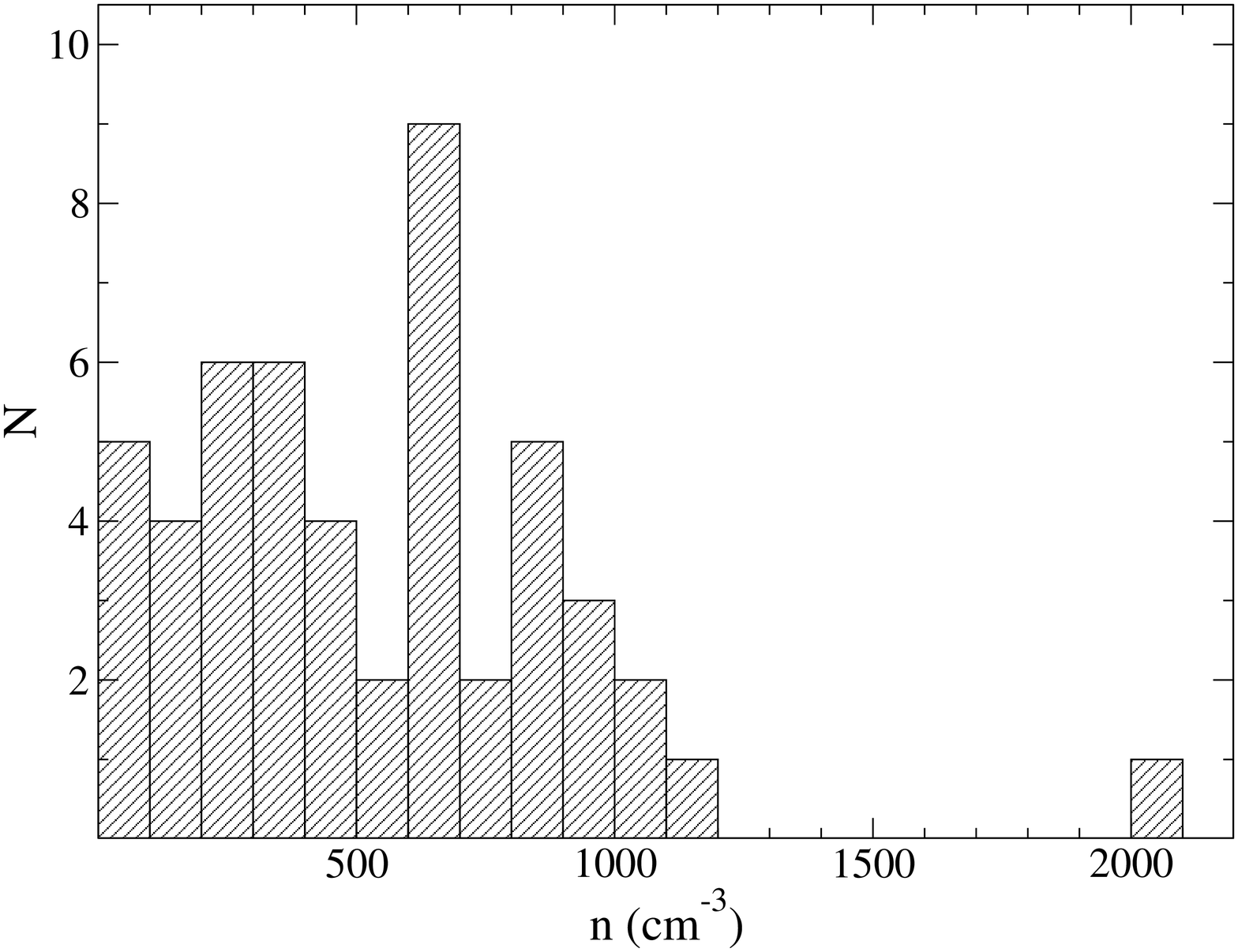}
\caption{Histogram containing electron density values (in $\rm cm^{-3}$) of Seyfert 2 galaxies
calculated using the  [\ion{S}{ii}]$\lambda6717/\lambda6731$ ratio line. The data were taken from
 \citet{kraemer94},  \citet{contini12},   \citet{koski78},    \citet{cohen83},  \citet{alloin92} 
\citet{enrique94}   \citet{radovich96},    \citet{osterbrock81},   \citet{rosa96},   \citet{osterbrock83}, \citet{phillips83},   \citet{goodrich83},
  \citet{shuder80},   \citet{durret88},   and \citet{shuder81}. }
\label{dens}
\end{figure}

\begin{enumerate}
\item Spectral energy distribution (SED):
 a two continuum components SED is assumed in the models. One 
 is the Big Bump component peaking at $\rm 1 \: Ryd$  with a high-energy and an infrared exponential cut-off, 
and the  other  represents the X-ray source that dominates at high energies. 
 This last component  is
characterized by a power law with a  spectral index $\alpha_{x}=-1$.  Its normalization was obtained taking into account the value $\alpha_{ox}=-1.4$ assumed for the optical to X-ray spectral index. 
Models assuming this  kind of SED  reproduce well a large sample of observational AGN data \cite[see][]{dors12}.

\item Ionization parameter $U$:  it is defined as $U={\rm Q_{ion}}/4\pi R^{2}_{\rm in} n  c$, where ${\rm Q_{ion}}$  
is the number of hydrogen ionizing photons emitted per second
by the ionizing source, $R_{\rm in}$  is  the distance from the ionization source to the inner surface
of the ionized gas cloud (in cm), $n$ is the  particle
density (in $\rm cm^{-3}$), and $c$ is the speed of light.   
The $U$ value was used as one of the  input parameters, therefore,  ${\rm Q_{ion}}$ 
and $R_{\rm in}$ are indirectly defined in each model. Cloudy changes the ${\rm Q_{ion}}$ value
when $R_{\rm in}$ is varied for  fixed $U$ and $n$ values, that results in the same
local cloud properties, yielding homologous models   with
the same predicted emission-line intensities \citep{bresolin99}. 
 We computed a sequence of models  with  $\log U$ ranging  from $-1.0$ to $-3.0$ (using a bin size of
0.5 dex).

To obtain a representative electron density value for NLRs of AGNs, we
compiled from the literature observational intensities of the line ratio of the
sulfur  [\ion{S}{ii}]$\lambda$6717/[\ion{S}{ii}]$\lambda$6731 
of 53 Seyfert 2 galaxies. Then,  we  computed  the electron density value $n$   for each object using the {\sc temden} routine of the nebular package of 
{\sc iraf}\footnote{Image Reduction and Analysis Facility, distributed by NOAO, operated by AURA, Inc., under
agreement with NSF.} assuming an electron temperature of 10\,000 K. In Fig.~\ref{dens} a histogram
of the obtained electron density values is shown. We can see that, for
most of the objects, $n$ is lower than 
about 1200 $\rm cm^{-3}$. The average of these values $<n>=537\:\rm \:
cm^{-3}$ was obtained and considered in our models.  This value is in consonance with the
densities derived  by \citet{bennert06}, who used high-sensitivity
spatially-resolved optical spectroscopy of a sample of Seyfert-2 galaxies.
   
\item Metallicity $Z$:  the metallicity of the gas phase in the models was linearly scaled to the solar metal composition with the exception of the N 
abundance,   which was taken from the relation  between N/O and O/H given by \citet{dopita00}. The C/O ratio was considered to be the solar 
value  $\rm \log(C/O)=-0.52$.   In the Cloudy code (version 08.00), the value  12+log(O/H)=8.69  taken from
\citet{allende-prieto01} is assumed as the  solar metallicity.    The metallicity  
range $-2.0 \: \le \log(Z/Z_{\odot}) \le \:0.60$ was considered in the models.  
For  models with $\log(Z/Z_{\odot}) =0.60$ and  $\log U$=$-2.5$, $-3.0$, the predicted
intensities of  \ion{C}{iv}$\lambda$1549   and/or  \ion{C}{iii}]$\lambda$1909  were about equal to zero and they were not consider
in our analysis.

\end{enumerate}

We  included internal dust in our models and not match with the observational data 
was possible, therefore, all models considered in this work are dust free. 
This result is in agreement with the one derived by \citet{nagao06a}, who showed that dusty models can not 
explain large observed values of the \ion{C}{iv}/\ion{He}{ii} line ratio  (see also \citealt{matsuoka09}). 
  The reason for models with dust can not explain the observed flux  of the lines considered is probably because
gas clouds in the high-ionization part of NLRs are dusty free, as suggested by 
\citet{nagao03}.

\begin{table} 
\centering 
\caption{Coefficients of the fitting of $\log (Z/Z_{\odot})=\rm a \times C43^{2} + b\times C43 + c$ for different values 
  of $\log U$.} 
\label{tab3} 
\vspace{0.2cm} 
\begin{tabular}{@{}l  c c r  @{}} 
\hline 
$\log U$           &     a  &    b      &   c        \\ 
 \noalign{\smallskip} 
 \multicolumn{4}{|c|}{Upper branch}  \\ 
$-$1.0             & $-1.45(\pm0.15)$ &  $-0.25(\pm0.16)$& $0.67(\pm0.04)$       \\ 
$-$1.5             & $-0.59(\pm0.06)$ &  $-0.77(\pm0.03)$& $0.35(\pm0.01)$       \\ 
$-$2.0             & $-0.18(\pm0.02)$ &   $-0.71(\pm0.03)$  & $-0.06(\pm0.01)$    \\ 
$-$2.5             & $-0.22(\pm0.02)$   &  $-0.79(\pm0.03)$  &  $-0.38(\pm0.01)$     \\ 
$-$3.0             & $ -0.12(\pm0.02)$  &   $-0.71(\pm0.06)$  &  $-0.63(\pm0.03)$     \\ 
\hline 
\noalign{\smallskip} 
 \multicolumn{4}{|c|}{Lower branch}  \\ 
$-$1.0             &    $4.60(\pm0.85)$&     $-4.03(\pm1.17)$  &  $-1.06(\pm0.39)$   \\ 
$-$1.5             &    $4.90(\pm0.78)$&     $-2.53(\pm0.80)$  &  $-1.49(\pm0.19)$   \\ 
$-$2.0             &   $1.13(\pm0.24)$ &      $1.37(\pm0.08)$ &  $-1.70(\pm0.01)$   \\ 
$-$2.5             & $0.76(\pm0.18)$   &  $1.82(\pm0.10)$     &  $-1.03(\pm0.01)$    \\ 
$-$3.0             & $1.02(\pm0.29)$    &   $2.81(\pm0.39)$  &  $-0.01(\pm0.11)$ \\ 
\hline 
\end{tabular} 
\end{table}

\section{C43- A new metallicity tracer}
\label{calib}

\subsection{$Z$-C43 calibration}

 Several metallicity indicators have been proposed to estimate the metallicity 
  using strong emission-lines  from  the gas phase of objects 
 without a direct determination of an electron temperature.  The idea is
 basically to calibrate  abundances using ratios among the strongest (easily
 measured) available emission lines.   In the case of star-forming regions,
  the pioneer work  by \citet{pagel79} proposed the 
 optical  metallicity  indicator $R_{23}$ (see also \citealt{pilyugin12}). 
 In general, it is preferable to use a   line ratio lower dependent on other physical
parameters than on the metallicity, for example, a line ratio with a weak
dependence on the ionization parameter $U$.

\begin{figure}
\centering
\includegraphics[angle=-90,width=12cm]{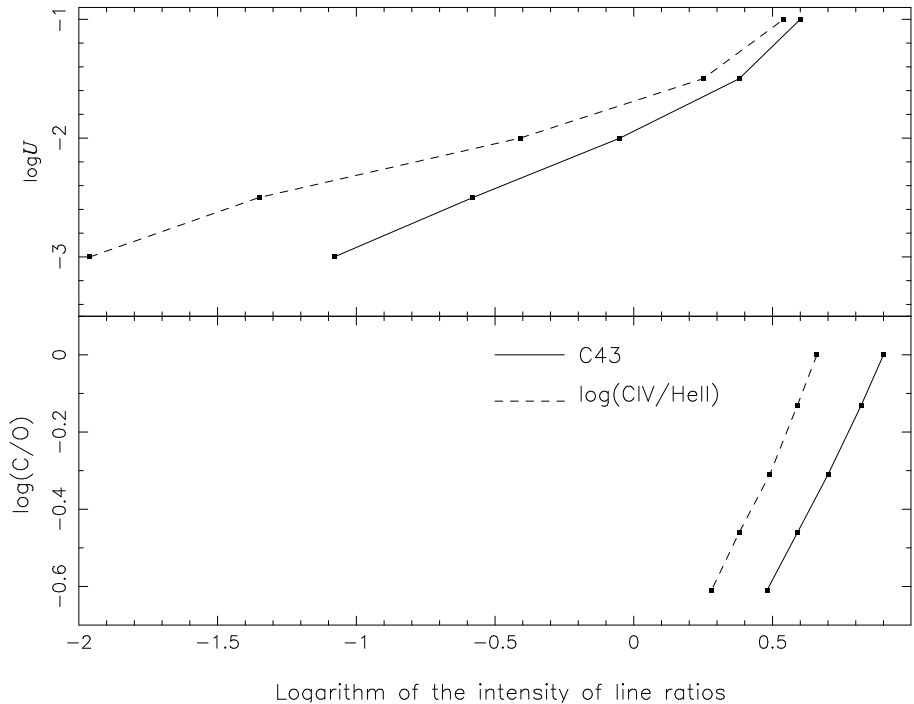}
\caption{Bottom panel: Abundance ratio of C/O versus the value of the metallicity indicators \ion{C}{iv}/\ion{He}{ii}  and C43 as indicated.
Lines connect the results of our models, represented by points, built
considering $\log U=-1.4$, and the other parameters as described in Sect.~\ref{calib}.
Top panel- Logarithm of the ionizing parameter versus the  value of the metallicity indicators \ion{C}{iv}/\ion{He}{ii}  and C43. A solar metallicity
was considered.}
\label{ana1}
\end{figure}

 For AGNs, metallicity indicators have also been
proposed along decades, for example, using strong optical narrow emission-lines 
(e.g. \citealt{thaisa98}) or UV-lines  (see \citealt{nagao06a} and references therein). The main difficult
in calibrating an index is that, in general, it depends on metallicity and 
other parameters, such as the ionization parameter, reddening corrections, electron gas density, abundance ratios (e.g. N/O, C/O; see \citealt{hamann99} for a review).  
 In particular, the \ion{C}{iv}/\ion{He}{ii} line ratio, suggested by \citet{nagao06a} as $Z$ indicator, is very dependent  on $U$, 
and a combination of this line ratio with another emission line from an ion with a lower ionization stage than $\rm C^{3+}$ can weakness this dependence.
In this sense, we proposed the use of the emission-line ratio  C43=log[(\ion{C}{iv}+\ion{C}{iii}])/\ion{He}{ii}] as metallicity indicator.
In Fig.~\ref{ana1} the  predicted variation of the C43 and \ion{C}{iv}/\ion{He}{ii}  for distinct values
of the   C/O abundance ratio and ionization parameters, obtained from our
models, are shown.  It can
be noted that, although the behavior of the C43 and the
\ion{C}{iv}/\ion{He}{ii} are very similar respect to
the C/O abundances (ranging the interval), a lower variation with the
ionization parameter is obtained for C43. 
The weak dependence of the C43 indicator with the ionization parameter becomes it in a more
reliable metallicity indicator than the \ion{C}{iv}/\ion{He}{ii}. This is
analogous to what is obtained in the optical wavelength range  for star-forming regions, where the $R_{23}$
parameter is less dependent on the ionization parameter than the
[\ion{O}{iii}]/H$\beta$ ratio \citep{kobulnicky99}.   The situation can be
 different in NLRs of AGNs than in  star-forming regions,  because  free electrons,
  neutral carbon and $\rm C^{+}$ ions (not considered in C43) 
  can co-exist in an X-ray Dominated Region  (see e.g.\ \citealt{mouri00}). 
 Therefore, the assumption that most of carbon is
in the form of $\rm C^{2+}$ or $\rm C^{3+}$  and that the metallicity can be estimated from the line ratio between these ions can be somewhat uncertain. 
However, even  taking this into account, C43 is more reliable than \ion{C}{iv}/\ion{He}{ii}, since
more than one ionization ion stage is considered,  tracing  a more   realistic assumption for the total abundance of C/H.

\begin{figure*}
\centering
\includegraphics[angle=-90,width=1.0\textwidth]{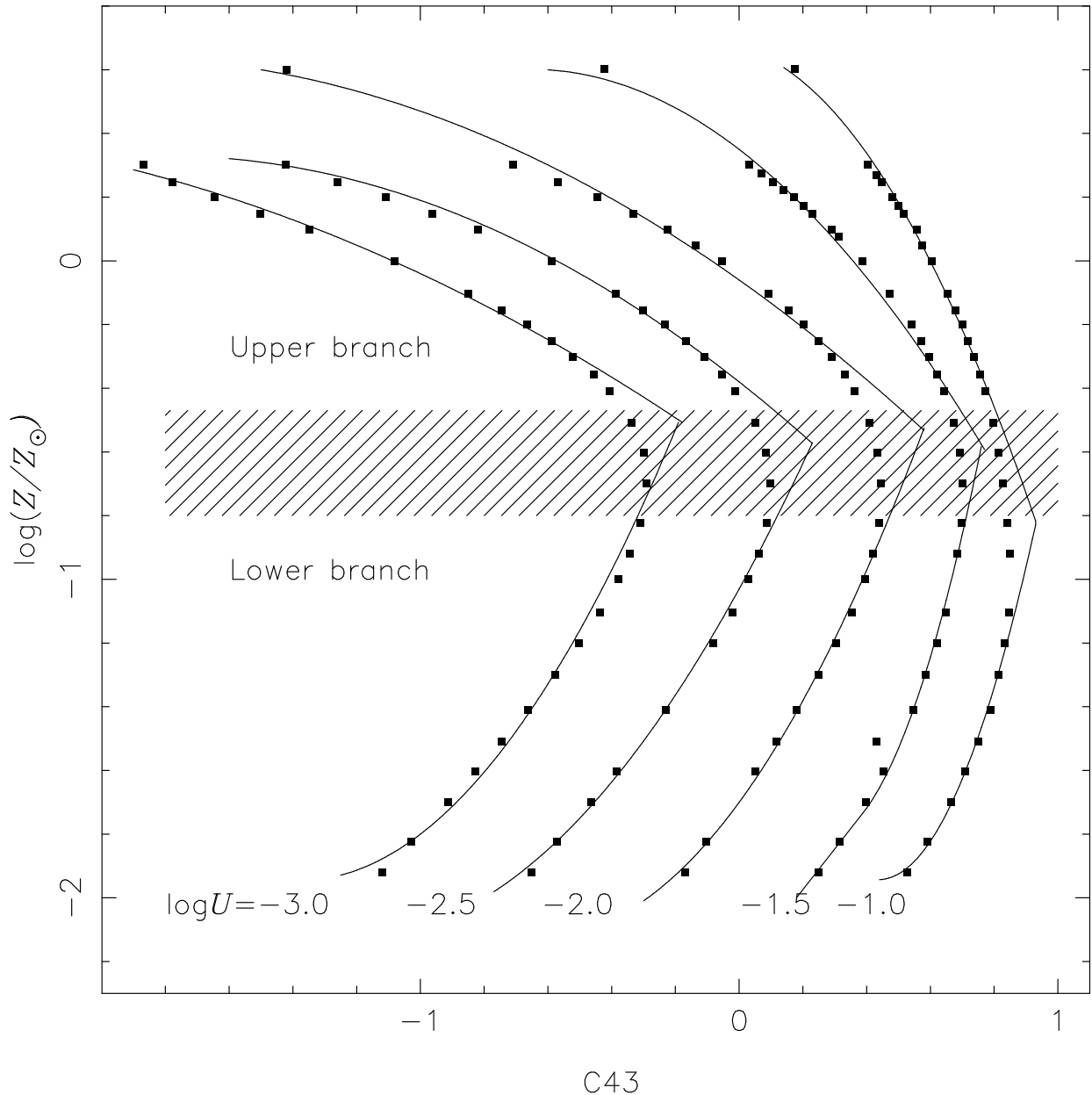} 
\caption{Logarithm of the metallicity in relation to the solar one vs.\ the
  C43. Curves  represent the  fits (see Table~\ref{tab3})  on model results
  (represented by points) with distinct ionization parameters  
  as indicated. The hatched area separates  the upper and lower branch as
  indicated.}
\label{ana2}
\end{figure*}

In Fig.~\ref{ana2} the  calibration between $Z$ and C43 considering different ionization parameter values  is shown.
 In Table~\ref{tab3}  the coefficients for second-order polynomial fits to the
models is given. We can see that C43 is double-valued with the metallicity,
yielding one branch to low metallicity (lower branch) and  other  to high metallicity (upper branch).  This problem is also found for other UV-line ratios (e.g. \ion{C}{iv}/\ion{He}{ii},
\ion{N}{iv}/\ion{He}{ii}) and  for the $R_{23}$ parameter (see \citealt{kewley08}). 
The inferred metallicities for AGNs,  even for the high redshift
ones, have been found to be solar or near solar (see e.g. \citealt{matsuoka09}), thus,
 hereafter we   only consider  the upper branch along the paper.

\begin{figure}
\centering
\includegraphics[angle=-90,width=12cm]{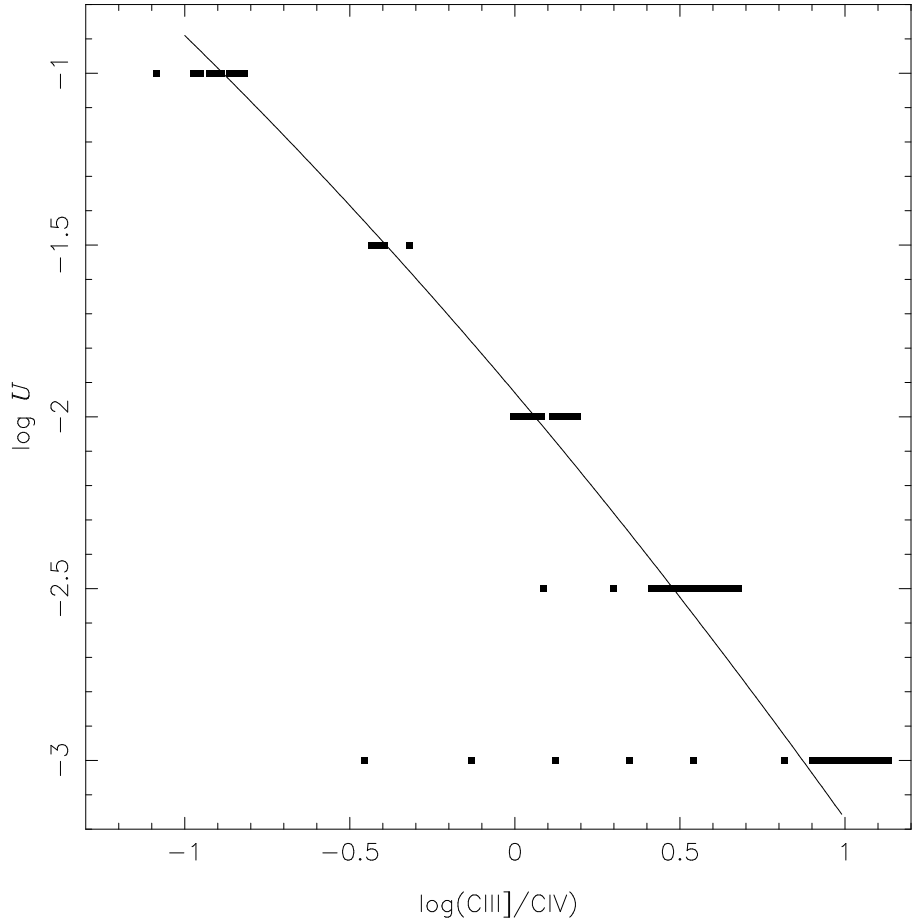}
\caption{Logarithm of the  ionization parameter versus log(CIII]/CIV). The points represent the results from our models
considering different values for $Z$ and $\log U$. The curve represents a fitting (Eq.~\ref{eq2}) to the average points for each  $\log U$ value.}
\label{ana3}
\end{figure}

The ionization parameter can be derived from the \ion{C}{iii}]/\ion{C}{iv}
  ratio \citep{nagao06a} which is weakly
dependent  on $Z$, mainly for high values of $U$.  In  Fig.~\ref{ana3}  we show this relation obtained from our models, which is represented by 

\begin{equation}
\label{eq2}
\log U=-0.10(\pm0.06)\times x^{2}-1.14(\pm0.02)\times x -1.93(\pm0.03),
\end{equation}
where $x$=log(CIII]/CIV). 

\subsection{Uncertainties in $Z$ estimations}

 Uncertainties in $Z$ estimations for  star-forming regions based on theoretical and/or empirical calibrations have been addressed for several authors.
 For example,  \citet{kewley08} showed that different  optical
methods or different empirical calibrations for the same  emission-line
ratios provide different oxygen  abundances (generally used as $Z$
tracer  of the gas phase),  
with discrepancies up to  a factor of 10. \citet{dors11}, who compared $Z$ estimations based on theoretical diagnostic diagrams
and on  direct  estimations of the electron temperatures, pointed out the importance of combining   two line ratios, one sensitive to
the metallicity and the other sensitive to the ionization parameter.  Regarding uncertainties in $Z$ estimations of AGNs based on UV-lines, 
few works have addressed this  subject.  In the case of the C43 index,
there are basically four sources of uncertainties,  which are discussed
  in what follows.
\begin{enumerate} 
\item C/O abundance ratio --- Since C43 index is dependent of the C/O  abundance, variations in this ratio
produce uncertainties in $Z$ determinations. We have performed a simple test to verify these uncertainties.  
 Considering the  averaged value for local AGNs (see Table~\ref{tab1}) 
 C43=0.53$\pm$0.09, and using the  $Z$-C43 calibration for $\log U=-1.5$ presented in Table~\ref{tab3},  we obtained $\log(Z/Z_{\odot})=-0.1$.
Now, if $ \rm \log(C/O)=-0.05$ is assumed to derive a new $Z$-C43 calibration (not shown),
  we derived $\log(Z/Z_{\odot})=0.20$.  Thus,
 a discrepancy by a  factor  $\sim$2 is obtained for $Z/Z_{\odot}$. 
\item Ionization parameter--- As seen in Fig.\ref{ana2}, the  $Z$-C43 calibration is dependent on
$U$. Using the fitting parameters shown in Table~\ref{tab3}  and considering that,
  according to the error in equation 1 and Table 2, $U$ can be
  estimated with an uncertainty up to 0.5 dex  (been about 0.1 for
    local AGNs), the $Z$ could ranges up to a factor of 3.   
\item Observational uncertainties--- Considering the observational uncertainty  of 
0.2 dex in the measured value of C43 and   the $Z$-C43 calibration for $\log(U)=-1.5$, we obtained that $\log(Z/Z_{\odot})$ ranges by about  a factor of 3.
\item Intrinsic uncertainty--- This uncertainty source is associated to the
  methods that use strong emission-lines 
to derive the metallicity. {\it Bona fide}  metallicity determinations for emission-line objects can only be  
achieved by estimations of  the electron temperature  ($T_{\rm e}$-method) of the gas phase (see \citealt{hagele08} and references therein).
Therefore, we  must compare the $Z$ values for our calibrations with
those  derived using the $T_{\rm e}$-method. Unfortunately, 
this was possible only for one object of our sample: NGC\,7674. Using
the optical data from \citet{kraemer94} and adopting the same procedure than \citet{dors11} we estimated $\log (Z/Z_{\odot})=-0.22$ 
for NGC\,7674  applying the  $T_{\rm e}$-method.  $\log U$ for this
object, calculated from Eq.~\ref{eq2}, is   about  $-1.7$.  Using the
correspondent fitting to the $Z$-C43 calibration (see Table~\ref{tab3}) we
estimated $\log (Z/Z_{\odot})=-0.28$, finding a difference of only 15 per cent  between these two estimations. We assumed this difference as
representative of the intrinsic uncertainty, even when more data are needed to
perform a confident statistical analysis of the influence of this uncertainty
on the $Z$-C43 calibration.
\end{enumerate}
Along this paper we consider the derived metallicity from C43 is correct by a
factor of  5 (about 0.7\,dex), the quadratic sum of the uncertainties
discussed above. This discrepancy would be smaller than that given by
\citet{kewley08} for the optical empirical parameters by a factor of 2.

\section{Results}
 \label{res}
 
\begin{figure*}
\centering
\includegraphics[angle=-90,width=1.0\textwidth]{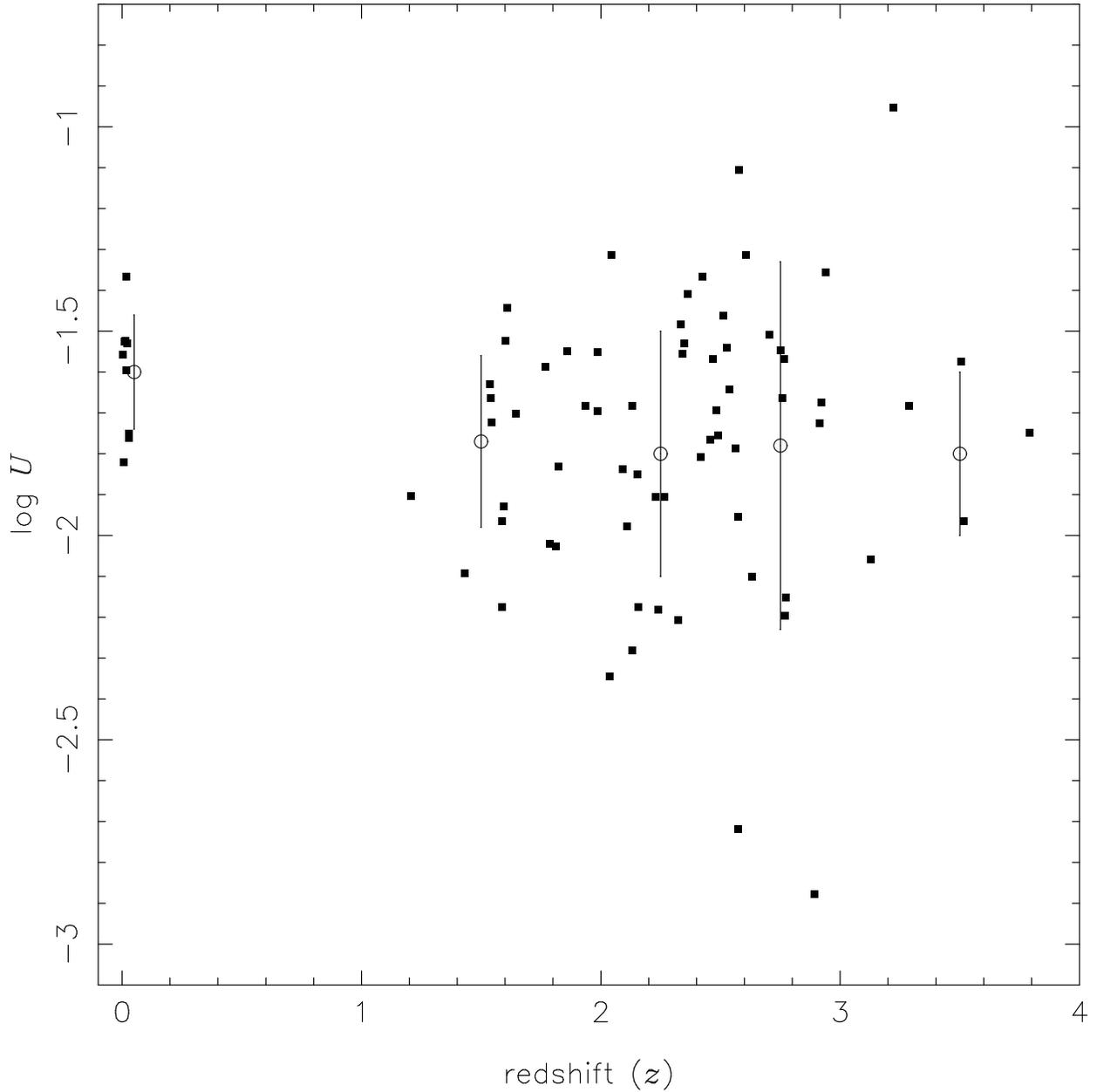} 
\caption{Logarithm of the ionization parameter vs.\ the redshift. Squares represent  $\log U$ values  obtained using Eq.~\ref{eq2} 
and the observational data presented in Table~\ref{tab0}. Circles represent the  average and their error bars the standard
deviation of $\log U$ for each redshift interval.}
\label{figion} 
\end{figure*}

 In Fig.~\ref{figion}  $\log U$ versus the redshift for the objects
in our sample, obtained using Eq.~\ref{eq2}, are plotted together with the
corresponding average and standard deviation for each redshift bin. We can
see  that the ionization parameters are in the range $ -2.8 \: \la \: \log U
\: \la  -1.0$, with an averaged value of about $-1.75\pm0.32$  dex.  This range is larger
than the one found by \citet{nagao06a}, who used the \ion{C}{iv}/\ion{He}{ii}
vs. \ion{C}{iii}]/\ion{C}{iv} diagnostic diagrams, finding $ -2.2 \: \la
\: \log U \: \la  -1.4$.

\begin{figure*}
\centering
\includegraphics[angle=-90,width=1.0\textwidth]{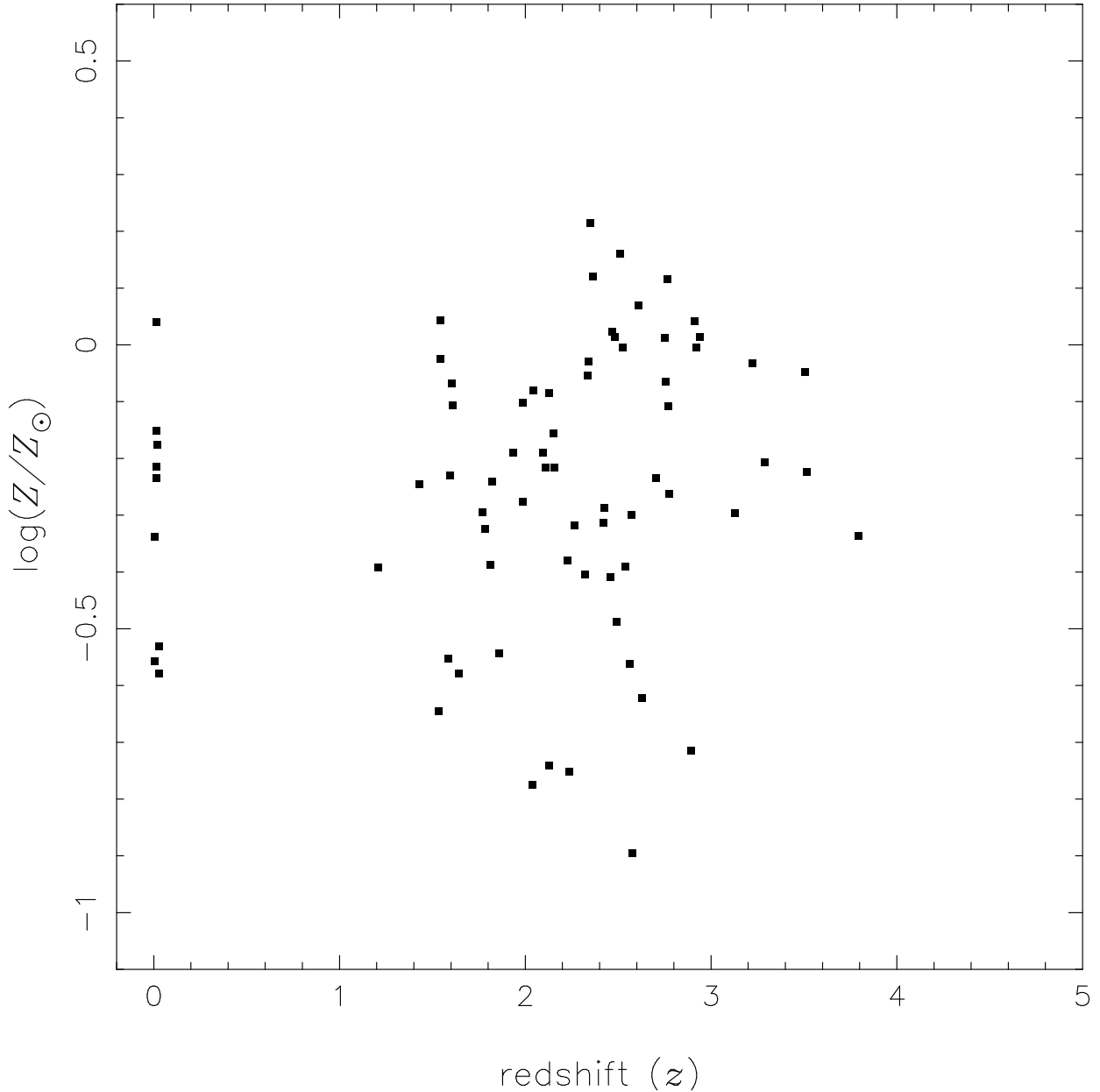} 
\caption{Logarithm of the metallicity in relation to the solar one vs.\ the
  redshift. Points represent estimations for the sample presented in
  Table~\ref{tab0} and considering the $Z$-C43 relations 
for different values of $\log U$  whose coefficients are given in
  Table~\ref{tab3}. The ionization parameter values were computed using
  Eq~\ref{eq2}.} 
\label{fig55}
\end{figure*}

\begin{figure*}
\centering
\includegraphics[angle=-90,width=1.0\textwidth]{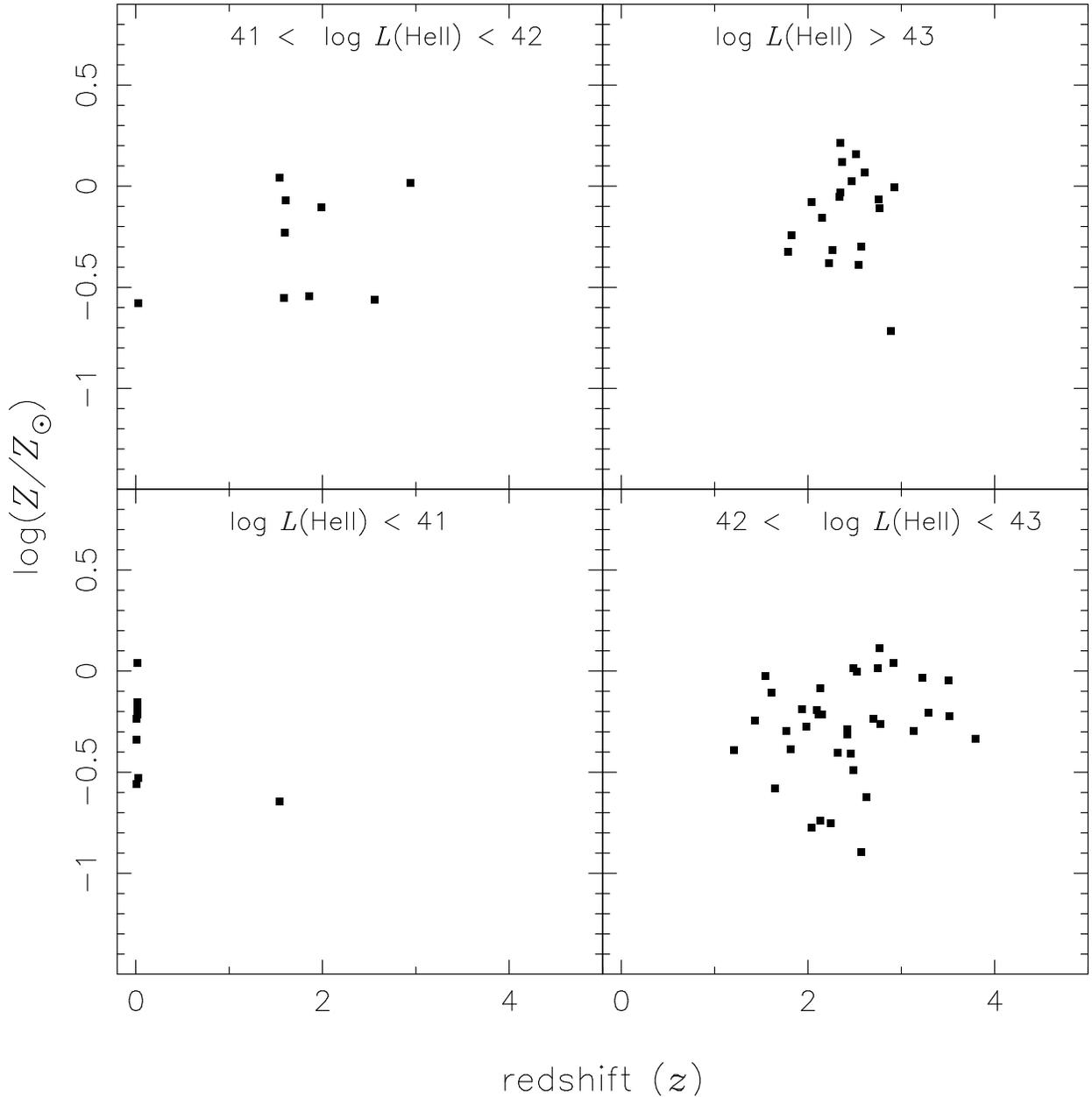} 
\caption{Such as Fig.~\ref{fig55} but considering different bins of luminosity
  as indicated in each plot.}
\label{fig56}
\end{figure*}

  To calculate the abundance for each object, we computed the ionization parameter using Eq.~\ref{eq2}
and  we selected the adequate set of coefficients for the $Z$-C43
calibration (see Table~\ref{tab3})  for the  closest available  $\log U$ value. 
In Fig.~\ref{fig55} the  logarithm of the derived metallicity in relation
to the solar one versus the redshift for the objects in our sample for which
were possible to estimate $U$ and $Z$ is presented. 
We can not note any metallicity decrease with the redshift.
 For some objects it was not possible to estimate $Z$ because 
some  emission-lines needed to calculate C43 were not available. Hence
the number of objects plotted in Figs.~\ref{fig55}  and \ref{fig56}
is smaller than the one
in Table~\ref{tab0} and Fig.~\ref{lumi}.

\begin{table}
\centering
\caption{Average metallicities for the objects in our sample
  considering different redshift and luminosity ranges. The number of objects 
  in each interval is given. The solar abundance of  $\rm 12+\log(O/H)=8.69$
  \citep{allende-prieto01} is assumed in the models.} 
\label{tab2}
\vspace{0.2cm}
\begin{tabular}{@{}ll  c  c@{}}
\hline                                   
      
 $z$             &                                                    &      $\log (Z/Z_{\odot})$         &     Number         \\    
 \noalign{\smallskip} 	
0-0.1          &     $\log \: L(\rm HeII)\:<\:41$          &           $-0.27(\pm0.19$)      &  8        \\
                  &     $41 < \log \: L(\rm HeII)\:<\:42$  &           $-0.57(\pm0.00$)     &   1         \\
                  &     $42 < \log \: L(\rm HeII)\:<\:43$  &           ---                            &  ---        \\
		   &     $\log \: L(\rm HeII)\:>\:43$         &           ---                           &  ---       \\
\noalign{\smallskip}     
1.0-2.0       &       $\log \: L(\rm HeII)\:<\:41$          &     $-0.64(\pm0.00$)          &  1          \\
                  &      $41 < \log \: L(\rm HeII)\:<\:42$  &      $-0.24(\pm0.25$)          &   6         \\
		   &     $42 < \log \: L(\rm HeII)\:<\:43$  &      $-0.27 (\pm0.16$)         &    9       \\
                   &     $\log \: L(\rm HeII)\:>\:43$         &       $-0.28 (\pm0.05$)         &   2       \\
\noalign{\smallskip}  		   
2.0-2.5       &       $\log \: L(\rm HeII)\:<\:41$          &       ---                              &  ---      \\
                  &      $41 < \log \: L(\rm HeII)\:<\:42$  &       ---                               & ---       \\
		   &     $42 < \log \: L(\rm HeII)\:<\:43$  &      $-0.37 (\pm0.25$)         &   13       \\
                   &     $\log \: L(\rm HeII)\:>\:43$         &       $-0.07 (\pm0.19$)         &     9       \\
\noalign{\smallskip}		   
2.5-3.0       &       $\log \: L(\rm HeII)\:<\:41$          &       ---                           &  ---       \\
                  &      $41 < \log \: L(\rm HeII)\:<\:42$  &       $-0.27(\pm0.40$)          &     2        \\
		   &     $42 < \log \: L(\rm HeII)\:<\:43$  &       $-0.23 (\pm0.35$)      &    8        \\
                   &     $\log \: L(\rm HeII)\:>\:43$         &        $-0.16(\pm0.28$)     &     8        \\	
\noalign{\smallskip}		   
3.0-4.0       &       $\log \: L(\rm HeII)\:<\:41$          &       ---                            &   ---         \\
                  &      $41 < \log \: L(\rm HeII)\:<\:42$  &        ---                           &  ---            \\
		   &     $42 < \log \: L(\rm HeII)\:<\:43$  &        $-0.19(\pm0.12$)      &   6           \\
                   &     $\log \: L(\rm HeII)\:>\:43$         &         ---                           &   ---            \\			   	                     
 \hline
\end{tabular}
\end{table}

\begin{figure*}
\centering
\includegraphics[angle=-90,width=1.0\textwidth]{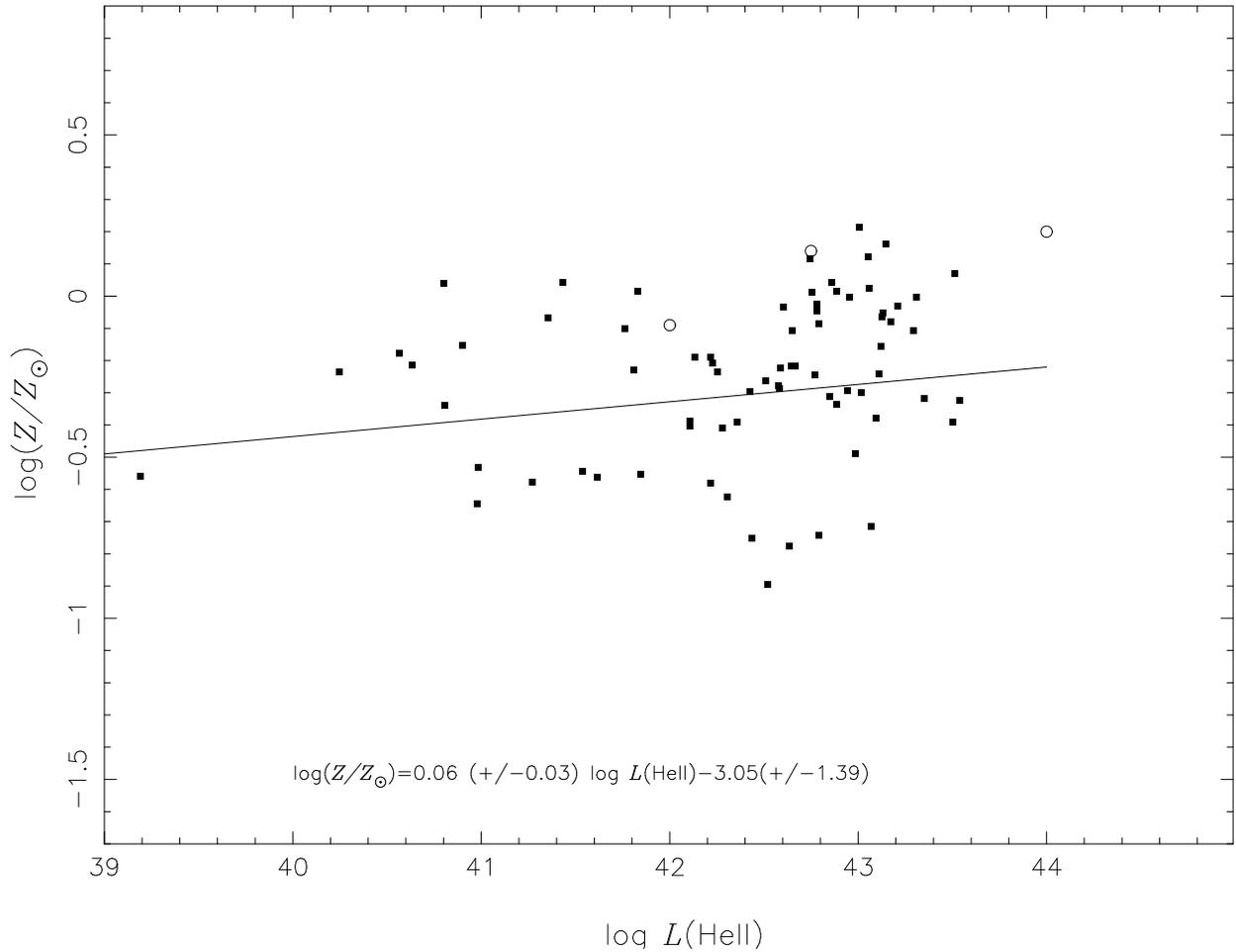} 
\caption{Logarithm of the metallicity in relation  to the solar one
    vs.\ the logarithm of the \ion{He}{ii} luminosity.  Squares represent 
estimations for our sample presented in Table~\ref{tab0} and considering the
$Z$-C43 relations.  
Circles represent mean values for HzRGs taken from \citet{matsuoka09}. 
A linear regression fit to the data is plotted.}
\label{fig57}
\end{figure*}

  In Fig.~\ref{fig56} the $Z$ estimations versus the redshift considering
different bins of luminosity is shown.   In Table~\ref{tab2} the $Z$ mean
values are given.  Although none $Z-z$ correlation can be noted,
objects with very low metallicity  ($\log(Z/Z_{\odot})\approx-0.8$), regardless of the
luminosity bin, are only found at redshifts  $1\: < \: z \: < \:3$.   In Fig.~\ref{fig57} the  
 metallicity versus the  \ion{He}{ii} luminosity  is presented.
 The mean values for HzRGs from \citet{matsuoka09} are also shown in
this plot. Although the large scatter of the points and the no so good linear
regression fit to our sample data, it seems to be a slight 
increase of $Z$ with the \ion{He}{ii} luminosity.

\section{Discussion}
\label{disc}

\begin{figure}
\centering
\includegraphics[angle=-90,width=16cm]{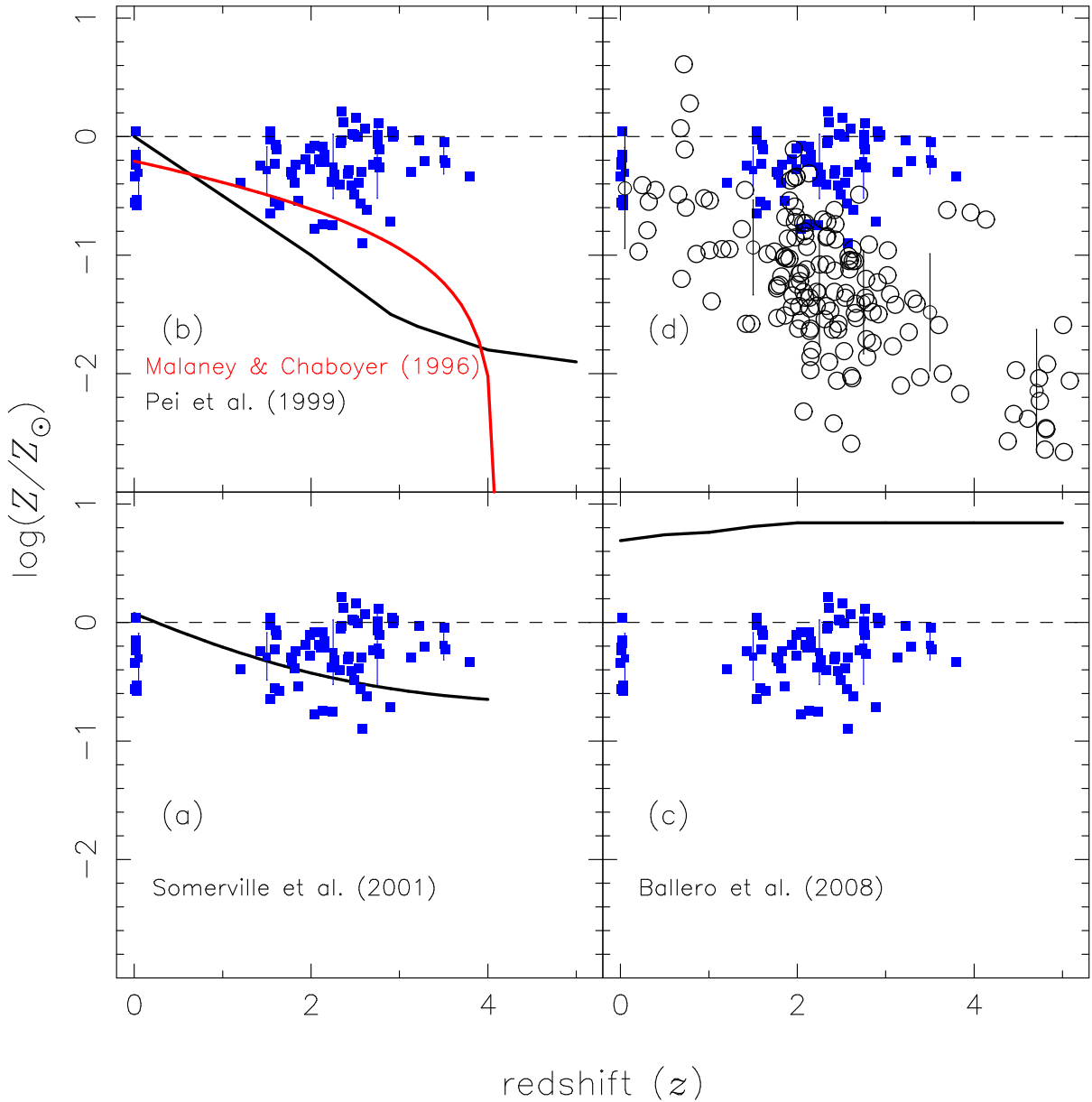}
\caption{Evolution of the logarithm of the metallicity in solar units $Z/Z_\odot$ with the redshift $z$. Squares without error 
bars represent our  metallicity results for AGNs and squares with error bars are the average of our metallicity results and their 
corresponding standard deviation considering different redshift intervals. In panels (a), (b) and (c),  curves represent prediction of cosmic chemical evolution models (see text).
In panel (d), circles represent metallicity estimations for Damped L$\alpha$ and sub-Damped L$\alpha$ galaxies 
via absorption lines by \citet{rafelski13}, \citet{fox07} and
\citet{kulkarni05}, and the circles with error bars represent their mean metallicity values for each redshift interval. Dashed lines represent the solar abundance value.}
\label{f5}
\end{figure}

  About two decades ago  the first determinations of metallicity in  high redshift star-forming galaxies \citep[$z\sim3$;][]{kobulnicky00}
 and  in  damped Lyman-$\alpha$ systems  \citep[$1.78\: < \: z\:< \:3$;][]{pettini94} were obtained. From these results, among others, a clear discrepancy arise: luminous
 high redshift galaxies are more metallic than DLAs at  the same redshift \citep{erb10b}. Likewise, the metallicity-redshift
 relation  followed by DLAs seems to be in consonance with some cosmic chemical evolution models that predict 
 a $Z$ increment with time (see e.g.\ \citealt{kulkarni13}). This kind of behavior has not been derived for using estimations of 
 $Z$ for AGNs. With the aim of compare our results with cosmic chemical model predictions and $Z$ determinations for other objects,
  we plotted them in Fig.~\ref{f5} as a function of the redshift. In what follows we briefly described the cosmic chemical models shown in this Figure. 
\begin{enumerate}
\item \citet{malaney96}--- Using the redshift evolution of the neutral hydrogen density inferred from observations of DLAs,
these authors calculated the evolution of elemental abundances in the Universe based on an analytical model.
From this work, models with a mean metallicity value (not corrected for dust obscuration) in a given redshift were considered.

\item  \citet{pei99}--- These authors obtained solutions for the  cosmic histories of stars, interstellar gas, heavy elements, dust, and radiation from stars and dust in galaxies using the available data
from quasar absorption-line surveys, optical imaging and redshift surveys, and the COBE DIRBE and FIRAS extragalactic infrared background measurements.
We considered   the mean metallicity  of interstellar gas in galaxies predicted by the best models from \citet{pei99}.
  
\item  \citet{somerville01}--- They  investigated several scenarios for the nature of the high-redshift Lyman-break galaxies using semi-analytic models of galaxy formation set within the cold dark matter 
merging hierarchy. 
From the models proposed by these authors, we considered the predictions for the average metallicity of the entire Universe  (taken from their Fig.~14), i.e. the total mass in metals divided by total mass of gas. 
This is the average  between the metallicities of the cold gas, stars, hot gas, and diffuse gas.

\item \citet{ballero08}-- These authors computed chemical evolution of spiral bulges hosting Seyfert nuclei, 
based on chemical and spectro-photometrical evolution models for the bulge of our Galaxy. We considered the metallicities predicted by those models 
built assuming a mass of the bulge of $2\: \times\: 10^{10}\:M_{\odot}$.
 \end{enumerate}

From Fig.\ref{f5} it can be seen that, for $z \la 3$ and considering the standard deviations, our metallicity estimations are in 
agreement with the predictions of the cosmic evolution models by \citet{somerville01}.  This agreement  
confirms the robustness of our $Z$ determinations using the C43 parameter. It also supports the \citet{somerville01} assumptions 
of a hierarchy galaxy formation and the form of the global star formation rate as a function of the redshift.  The independence of the metallicity with the redshift
derived from our results can be  biased  by an observational constrain in the way that we are using only the data of luminous objects at high redshift (see Fig.~\ref{lumi}),
i.e. at such redshifts we are able to observe only the most metallic objects. 
 For $z \:>\: 3$, we have few Z determinations and there could be incompleteness effects in the sample. Therefore, definite conclusions can not be obtained
for this redshift range.

Models by \citet{malaney96} and  \citet{pei99} predict higher metallicities than our estimations (see Fig.~\ref{f5}). 
This could be due to the \ion{H}{I} density values used as input in the models of these authors rather than an incorrect
 selection of the star formation parameters, which control the enrichment of the ISM.  The highest discrepancy is found for the 
 model evolution by \citet{ballero08}, which shows higher values of $Z$ than the ones derived by us. Interestingly, the results from 
 all these chemical evolution models inferred a solar metallicity for the Local Universe, except for the one by \citet{ballero08}.

In panel (d) of Fig.~\ref{f5}, we compare the cosmological mean metallicity ($<Z>$) computed  for
individual elements (e.g. Zn, S and Si) of DLAs and sub-DLAs (taken from
 \citealt{rafelski13}, \citealt{fox07}  and \citealt{kulkarni05})  with our metallicity estimations. The abundance solar value is also indicated in this plot. 
Our results  predict  a mean metallicity for
local objects in agreement with the solar value (12+log(O/H)=8.69).  This
value is about the same that the maximum oxygen abundance derived for the
central parts of spiral galaxies  \citep{pilyugin07}, and for circumnuclear
star-forming regions in both AGNs \citep{dors08} and normal galaxies
\citep{diaz07}. 
 Concerning the  $<Z>$ in DLAs and sub-DLAs, they tend to decrease with the redshift while our estimations for 
AGNs present an almost flat behavior, showing an agreement only in the Local Universe. 
\citet{somerville01} pointed out 
that $<Z>$ estimations in DLAs can be systematically underestimated 
due to two factors. First,  dusty high metallicity systems might dim quasars in the line
of sight \citep{pei95}. Second,   the outermost 
regions  of spiral galaxies have often lower $Z$ than central regions, thus,
$Z$ estimations  of objects at high redshift, not spatially resolved, represent
values lower than the one attributed to the active nuclei. 
The  $Z$ estimations for the objects in our sample are affected  at least by the second
factor. Therefore, it is unlikely that the discrepancy 
found in Fig.~\ref{f5}(d) may be due to the factors discussed by Somerville and collaborators.

As can be seen in Fig.~\ref{f5}, we found no clear metallicity evolution
with the redshift.  Similar
result was also found by  \citet{matsuoka09} and \citet{nagao06a}. 
It is worth to emphasize that, independently of the luminosity (see
Fig.~\ref{fig56}), very low metallicity $Z/Z_{\odot}\approx -0.8$ is found for
some AGNs in the range $1.5 \:< \:z \:< \:3$,  in
consonance with the $<Z>$  found in DLAs and sub-DLAs.   Except for
the local objects, the mean abundance value estimated by us using the $Z$-C43
calibration is higher than the mean value for DLAs and sub-DLAs for each
redshift interval.   
In fact, \citet{nagao06a} presented two interpretations from their analysis:
 (i) the narrow line regions of AGNs have sub-solar metallicities ($-0.7
\: \la \: \log(Z/Z_{\odot}) \: \la  \:0$) if low-density gas clouds 
with $n \: \la \:10^{3} \: \rm cm^{-3}$ are considered in their
photoionization models;  (ii) a wider range of gas metallicity
($-0.7 \: \la \: \log(Z/Z_{\odot}) \: \la  \:0.7$) for high-density gas clouds
with $n \: \approx \:10^{5} \: \rm cm^{-3}$. 
Although,  in some cases (see e.g. \citealt{peterson13}), high values of
electron density (in the order of $10^{5} \: \rm cm^{-3}$) were derived
for NLRs, we showed that densities of $\sim 500\:\rm \: cm^{-3}$ 
are representative for AGNs. This low densities yield that very low
  metallicity be derived for some objects at high redshift.

 \section{Conclusions}
 \label{conc}
 
We proposed here a metallicity indicator based on the emission-line
ratio C43=(\ion{C}{iv}+ \ion{C}{iii}])/\ion{He}{ii}. This index seems to
be  a more reliable metallicity indicator than other proposed in the
literature since it has a weak dependence on the ionization parameter.
We confirmed the no metallicity evolution of NLRs with the redshift that
was pointed out by previous works. Our results predict a mean metallicity for
local objects in
agreement with the solar value (12+log(O/H)=8.69). This mean value is also
in consonance with the maximum oxygen abundance derived for the central
parts of spiral galaxies.
 For $z \la 3$ and considering the standard deviations, our metallicity
estimations through the C43 parameter are in agreement with the predictions
of the cosmic evolution models by \citet{somerville01}.  
For $z \:>\: 3$, we have few Z determinations and there could be
incompleteness effects in the sample produced by the observational constrain of having data
 only from the most luminous objects. Therefore, the sample of objects with $z
 \:>\: 3$ is needed to be enlarged, mainly for brightness objects, to avoid
 possible observational biases and to improved the
conclusions about the metallicity evolution of AGNs with the redshift.

\section*{Acknowledgments}
We are very grateful to the anonymous referee for his/her complete and
  deep revision of our manuscript, and very useful comments and
suggestions that helped us to substantially clarify and improve our work.    We thank   Gary
Ferland for providing the  photoionization code Cloudy to the public.
OLD and ACK are grateful to the FAPESP for support under
grant 2009/14787-7 and 2010/01490-3, respectively.  MVC and GFH thank the
  hospitality of the Universidade do Vale do Para\'iba. OLD thanks the
  hospitality of the University of Heidelberg where part of this work was done.

\label{lastpage}

\end{document}